# Polarization Symmetry Breaking in Nitrogen under High Pressure

Meifang Pu and Li Lei*

*Institute of Atomic and Molecular Physics, Sichuan University, Chengdu 610065, China*

*Corresponding author: lei@scu.edu.cn

An order parameter with broken polarization symmetry is proposed for the pressure-induced molecular dissociation transitions in nitrogen. The experimental dissociative transition pressure dependence of the calculated order parameter is well described by a power law of the form ***P=kq²***. Our results unveil the important role of symmetry breaking in nitrogen under high pressure and further demonstrate a path from dense molecular fluid to fully polymeric phases by breaking local polarization reversal symmetry.

Symmetry breaking, first introduced by Landau, has spread widely in many branches of physics. When a system approaches a continuous phase transition, only the initial symmetry and how it changes as a result of the transition are important. An order parameter, which is a measure of the degree of symmetry breaking across the boundaries in phase transition, continuously grows starting from zero to nonzero [1]. High-pressure phase transition involves a symmetry breaking process in the same way. Pressure-induced dissociative transitions in prototypical linear diatomic molecules, such as $H_2$ and $N_2$ [2-4], lead to rich solid polymorphous. Such important high-pressure phase transitions, however, are a lack of the description of symmetry breaking.

In this Letter, we introduce an order parameter with broken polarization symmetry in conjunction with Landau phase transition theory to summary the pressure-induced dissociation transitions in nitrogen. We demonstrate a path from dense molecular fluid to fully polymeric phases by breaking local polarization reversal symmetry and further unveil the important role of symmetry breaking in nitrogen under high pressure. As the pressure is raised on nitrogen molecules, intermolecular interaction increase, polarization density change, and a variety of solid molecular phases are formed by pressure-induced intermolecular-dissociation transitions [4-7]. There have been found at least eight different molecular solids ($\beta$, $\delta$, $\delta_{loc}$, $\varepsilon$, $\zeta$, $\zeta'$, $\kappa$ and $\eta$) by compressing fluid nitrogen up to 130 GPa at room temperature [8-15]. Intermolecular structural characteristics make solid molecular phases differ from each other. Solid molecular nitrogen retain N≡N triple bonds up to at least 130 GPa, exhibiting broad low-frequency lattice phonon modes (100~1000 $cm^{-1}$) and sharp high-frequency stretching vibron modes (2300~2500 $cm^{-1}$) [16]. Above 150 GPa, the nitrogen becomes dark red and exhibits single-bonded (N-N) nonmolecular characteristics accompanied by the marked decreases in Raman intensity [11,16] and electrical resistance [17]. Upon heating at high pressures, the triple-bonded solid molecular nitrogen dissociates to form a single-bonded cubic gauche phase (cg-N) at above 110 GPa [18-20], or a layered-polymeric phase (LP-N) at above 150 GPa [21]. Nitrogen exhibits complicated phase behaviors under high pressure [Fig. 1(a)]. However, the quantitative description of dissociative transitions in terms of broken symmetry has not been done, and the phase evolution pathway is lack of a unified physical picture.

In order to open an avenue for broken symmetry analysis of pressure-induced prototypical dissociative transitions in nitrogen, we carefully determine the dissociative transition pressures by the discontinuities in high-frequency vibrons with high-pressure Raman spectroscopy using a 532 nm excitation wavelength (see Supplemental Material [22]), perform local polarization symmetric analysis for different solid molecular phases, and parametrize the symmetry breaking by a new order parameter with respect to linear molecular polarization. Assuming that the Raman polarization is produced by the local field associated with N≡N symmetrical linear stretching vibration, N≡N molecular pairs in a different local field exhibit different high-frequency vibron behaviors. Pressure-induced polarization symmetry breaking leads to the splitting of high-frequency Raman vibrons [Fig. 1(b)]. Nitrogen is unable to retain N≡N triple bonds upon heating at high pressures. We synthesized the polymeric cg-N at 111 GPa and 2000 K or 134 GPa and 1500 K [Fig. 1(a)] without use of any irradiation absorbing agents in a custom-built double-sides laser-heated diamond anvil cell (LHDAC) system with a 1064 nm fiber laser. We use ruby luminescence [23], the high-frequency edge of the diamond phonon [24-25], and the calibrated $v_1$ vibrational frequency [7, 26] as the pressure scale. After further compression of the cg-N to 157 GPa at room temperature, the cg-N was transformed to the LP-N by direct laser heating up to 2000 K at 152 GPa [Fig. 1(c)], suggesting a direct transformation between two polymeric phases. About 2-5 GPa pressures were drop after the laser heatings. Laser heating temperatures were either measured by spectroradiometric method or estimated. Fig. 1(c) shows Raman spectra of different solid phases of nitrogen, the disappearance of high-frequency vibron modes suggest the violation of N≡N linear stretching vibration, and the appearance of low-frequency of lattice modes at 800-1200 cm$^{-1}$ clearly indicate the formation of a non-molecular solid with N-N single bonded.

The end of the intermolecular-dissociation processes correlate with the onset of the intramolecular-dissociation transition. There is a path from fully disordered fluid nitrogen to a fully ordered polymeric phase by breaking structural and polarization reversal symmetry. To specify the disorder (high symmetry) - order (low symmetry) phase transitions in nitrogen quantitatively, we can define an order parameter, $q$, by

$$q = \left(\frac{V}{N}\right)^{-\frac{1}{3}} \cdot \frac{1}{n}\sum_{i=1}^{n}\frac{C_i}{M_i} \qquad (1)$$

where $C_i$ is the number of local molecular cluster with same structural symmetry (labeled $i$) in the unit cell, and $M_i$ is the number of N≡N linear molecular pairs with same local polarization per unit cell. $V$, $N$, and $n$ are the volume of unit cell in Å³ at the transition pressure, the number of triple-bonded nitrogen atoms per unit cell, and the number of Raman stretching vibration modes for each phase, respectively. Note that nitrogen molecular pairs with same local polarization are all in the same type of local molecular clusters. If we define a mean phonon polarization degeneracy number $D = \frac{1}{n}\sum_{i=1}^{n}\frac{C_i}{M_i}$ and a mean polarization length $L=(V/N)^{1/3}$ describing one-dimensional average length for a nitrogen atom, then

$$q = D/L \qquad (2)$$

here $q$ indicates the phonon polarization degeneracy number on unit length. Obviously, $L$ has dimension **L**, and $D$ is a dimensionless quantity. Thus, $q$ has dimension $\mathbf{L^{-1}}$.

In the case of fully disordered fluid nitrogen, there is only one N≡N symmetrical stretching mode ($\nu_1$) associated with local molecular polarization, which is homogenous disordered in the whole space. Therefore, $n$ and $C_1$ for the fluid nitrogen are 1 and 0, respectively. In addition, $M_1$, $N$ and $V$ are all ∞. From the Eq. (1), the order parameter for the fluid nitrogen $q^{LN}=0$. $\beta$-N is the first high-pressure phase of nitrogen at room temperature. The hexagonal $\beta$-N with space group $P6_3/mmc$ [27] possesses only one high-frequency Raman vibron ($\nu_1$) with regard to N≡N stretching vibrational mode in the local molecular clusters, in which six diatomic molecules interact with *Van der Waals* forces to form a symmetric clusters [Fig. 2(a)]. Each N≡N stretching vibration ($\nu_1$) is completely equivalent to the local environment in the molecular cluster. Due to there are two equivalent local molecular clusters in the unit cell, the order parameter for $\beta$-N is $q^\beta= 0.12$ with $L=1.397$ (Å) and $D =1/6$ (see

Supplemental Material [22]).

$\delta$-N is transformed from $\beta$-N at about 5 GPa with the $\nu_1$ mode splitting into two modes, $\nu_1$ and $\nu_2$ [Fig. 1(b)]. $\delta$-N has lower symmetry than $\beta$-N. As shown in Fig. 2(b), four N≡N pairs in the unit cell interact with *Van der Waals* forces to form two types of local molecular clusters (a cubic-like and a disk-like), which correspond to the $\nu_1$ and $\nu_2$ modes, respectively. Accordingly, $q^\delta$=0.162 with $L$=1.541 (Å) and $D$=1/4 [22]. The $\delta$-$\delta_{oc}$ phase transition is identified at 8.9 GPa by the discontinuities of $\nu_1$ and $\nu_2$ vibrons (see Supplemental Material [22]). $\delta$-N and $\delta_{oc}$-N are stable at 5-17.4 GPa at ambient temperature, which are in good agreement with the previous study [15]. Above 17.4 GPa, $\delta_{oc}$-N transforms to $\varepsilon$-N with the $\nu_2$ further branched out into $\nu_{2c}$ and $\nu_{2a}$ [Fig. 1(b)]. We find that the rhombohedral $\varepsilon$-N with space group $R$-$3c$ contains three types of local molecular clusters [Fig. 2(c)], and the three vibrational modes ($\nu_1$, $\nu_{2a}$, and $\nu_{2b}$) are associated with local symmetrical stretching vibrations in the linear chain nitrogen molecules ($\nu_1$), the counterclockwise rotation three-pair nitrogen molecular clusters ($\nu_{2a}$) and the clockwise rotation three-pair nitrogen molecular clusters ($\nu_{2c}$), respectively. Therefore, $q^\varepsilon$=0.25 with $L$=2.222 (Å), and $D$=5/9 (see Supplemental Material [22]).

We determine that the transition pressure for $\zeta$-N is about 56 GPa, at which the $\nu_{2c}$ further branches out into $\nu_{2b}$ and $\nu_{2c}$ [Fig. 1(b)]. The orthorhombic $\zeta$ phase with space group $P222_1$ contains four local symmetrical polarization modes ($\nu_1$, $\nu_{2a}$, $\nu_{2b}$, and $\nu_{2c}$) in each unit cell [Fig. 2(d)]. We have $q^\zeta$=0.517 with $L$=1.934 (Å), and $D$=1 (see Supplemental Material [22]). $D$=1 suggests that local molecular clusters are broken into relatively independent linear molecular pairs. The $\zeta$-N and its derivative $\zeta'$-N can be stable at 56-106 GPa [Fig. 1(b)], and the $\nu_{2c}$ mode for $\zeta'$-N started to turn around at about 87 GPa, which is ascribed to the slight rotation or distortion of molecular clusters in the lattice. Our results demonstrate that pressure-induced structural and polarization symmetry breaking leads to a series of sold molecular phases [Fig. 2(e)].

The $\kappa$-N with five vibrons ($\nu_{2a}$, $\nu_{2b}$, $\nu_{2c}$, $\nu'_{2a}$ and $\nu'_{2b}$) can be stable at 106-126 GPa [Fig. 1(b)], which is a rough agreement with the previous work by Gregoryanz *et*

*al.* using *in-situ* X-ray diffraction [28]. Above 126 GPa, the κ-N then transforms to an amorphous-like η-N with significant decrease in Raman intensity [Fig. 1(b)]. A small peak at 136 GPa in the [Fig. 1(c)] could be come from the residual of phase transitions. Because of the lack of detailed structural information about $δ_{loc}$, ζ', κ and η phases, the order parameters for these phases are absent. Due to the structural similarities between δ and $δ_{loc}$, as well as ζ and ζ', the order parameters of δ and ζ are considered to be almost identical to their derivative phases.

When triple-bonded solid molecular phases are dissociated into single-bonded polymeric phases (cg-N or LP-N), as shown in the [Fig. 1(c)], all the high-frequency vibron modes (2350-2500 cm$^{-1}$) disappear and meanwhile low-frequency lattice modes (800-1200 cm$^{-1}$) appear, suggesting the end of the intermolecular dissociations and the onset of the intramolecular dissociative transitions. For any given fully polymeric phase or amorphous nonmolecular phase, N≡N triple bonds cannot be retained, so that $q^{N\text{-}N}=1$ (see Supplemental Material [22]).

According to the discussion above, the calculated order parameters as the function of experimental transition pressures for nitrogen are plotted in Fig. 3. The room-temperature dissociative transition pressure, ***P***, dependence of the order parameter, ***q***, is well described by a power law of the form $q \sim P^α$. The fitting curves on this expression are also plotted in the Fig. 3. The nonmolecular nitrogen and LP-N are well on the extended line of the fitting curve based on the room-temperature transition pressures for the solid molecular phases (β, δ, ε, and ζ). The value of the exponent ***α*** is determined by linear regression from the slop of the *lnq* versus *lnP* curve (Fig. 3 inset), yielding a value of *α*≈0.5. So we have,

$$P=kq^2 \qquad (3)$$

where ***k*** is the dissociative coefficient, which possesses the same dimension of the physical quantity force; and $q^2$ has dimension $\mathbf{L}^{-2}$, which is correlated with the physical quantity area. For room temperature, the dissociative transition pressure is 150 GPa, then ***k*** has 1.5×10$^{-9}$ N. It is noted that the cg-N exhibits a deviation from the fitting line (Fig. 3). This is due to the fitting curve that are based on the room-temperature dissociative transition pressure for those solid molecular phases (β,

$\delta$, $\varepsilon$, and $\zeta$), which can well describe the dissociative transitions under room temperature.

The Eq (3) can be useful for inferring dissociative transition pressure for a given solid molecular phase. $\lambda$-N, for example, is a solid molecular phase (low symmetry) formed by compression of liquid nitrogen (high symmetry) at low temperature (77 K) up to >32 GPa [29]. According to the reported $P2_1/c$ structure for $\lambda$-N, it has $M_1$=1, $C_1$=1, $M_2$=1, $C_2$=1, $V$=39.85 (Å$^3$), $N$=4, $n$=2, $L$=2.151 (Å), and $D$=1. So $q^\lambda$=0.465. The transition pressure for $\lambda$-N can be inferred to be 32.4 GPa, which is very close to the experimental pressure 32 GPa [29]. Moreover, the relation in Eq (3) is also expected to be useful for the study of the pressure-induced dissociation transitions in other linear diatomic molecules such as H$_2$ and O$_2$. If the order parameters for the solid molecular phases were given, the intramolecular-dissociation transition pressures could be determined by extending the pressure dependence of the order parameter fitting curve on $q$=1.

In conclusion, we demonstrates a path from dense molecular fluid nitrogen to fully polymeric phases by breaking structural and polarization reversal symmetry, and we introduce an order parameter, $q$, with a broken polarization symmetry in conjunction with Landau phase transition theory to summary the pressure-induced dissociative behaviors in nitrogen. The single-bonded nitrogen are well on the extended line of the $P=kq^2$ fitting curves.

We acknowledge support by the National Natural Science Foundation of China (Grant No. 11774247).

**Figure caption**

Fig. 1(a) P-T phase diagram for nitrogen and experimental P-T routes. The solid lines are thermodynamic boundaries and the dashed lines signify the kinetic lines. The blue dashed lines show the compression paths at room temperature, and the red dashed lines show laser heating paths which will produce cg-N and LP-N in this work. The open square (cg-N), open diamond (cg-N) and regular open triangle (LP-N) are from Ref. [18], Ref [26] and Ref. [21], respectively. The open circle (nonmolecular-N) is from Ref [17]. The solid squares (cg-N) and inverted solid triangle (LP-N) are the experimental data in this work. (b) Pressure dependence of high-frequency Raman vibrons for nitrogen. (c) Raman spectra of different solid phase of nitrogen.

Fig.2 (a)-(d) Left panel: crystal structures for $\beta$, $\delta$, $\varepsilon$, $\zeta$-phases. Right panel: Raman vibrons with different local symmetrical stretching vibrations. (e) Pressure-induced structural and polarization symmetry breaking leads to a series of sold molecular phase transitions.

Fig. 3. The calculated order parameters as the function of experimental transition pressures for nitrogen. The blue solid line is the power-law fit ($P=kq^{\alpha}$) to the room-temperature dissociative transition pressure data. The solid circles are experimental data from our Raman measurements. The open circle (nonmolecular-N), open square (cg-N), open diamond ($\lambda$-N) and regular open triangle (LP-N) are from Ref. [17], Ref. [18], Ref. [29] and Ref. [21], respectively. The solid squares (cg-N) and inverted solid triangle (LP-N) are the data in this work. Inset: The *lnq* versus *lnP*. The red solid line in the inset is the linear fit to the data, yielding $\alpha\approx0.5$.

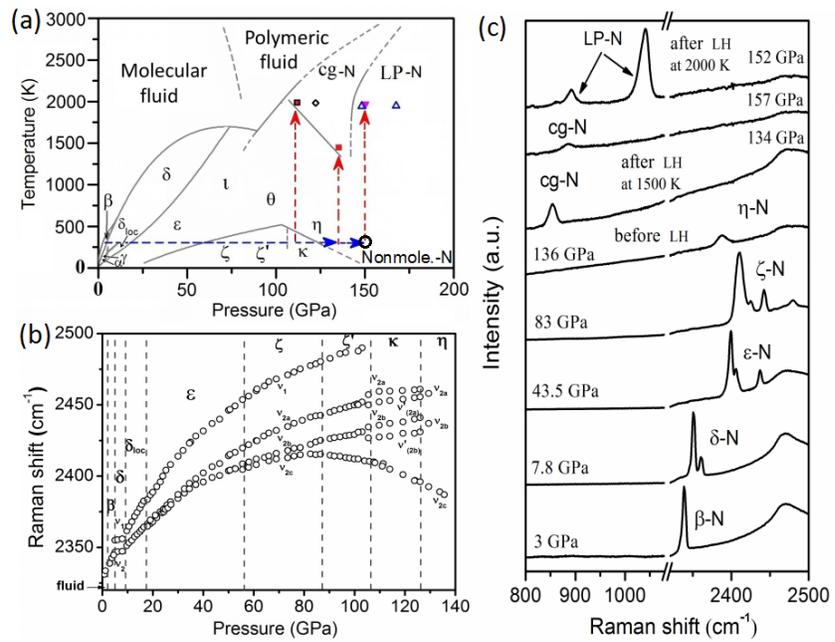

Figure 1

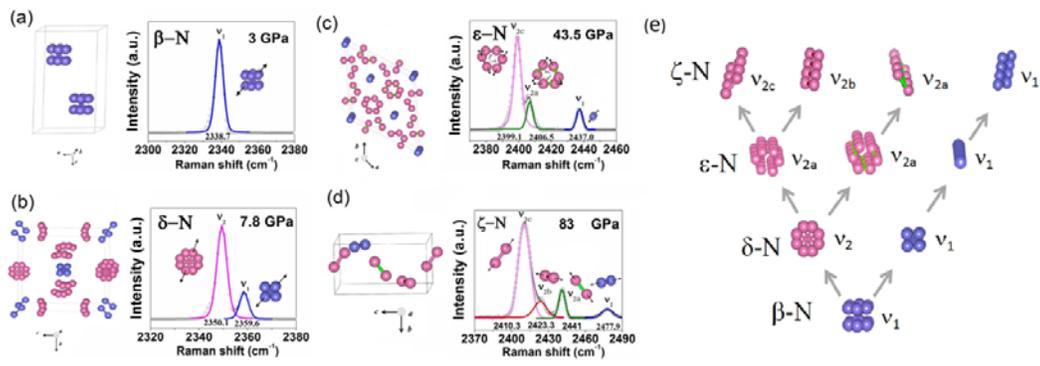

Figure 2

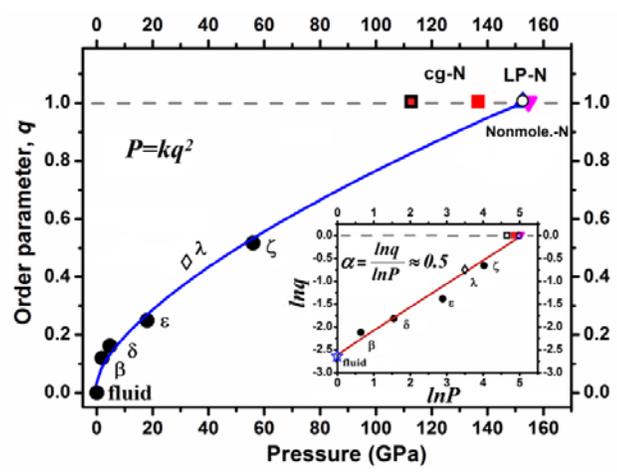

Figure 3